\documentclass[12pt,preprint]{aastex}

\usepackage{natbib}

\def\spose#1{\hbox to 0pt{#1\h\> \> SS}}
\def\simlt{\mathrel{\spose{\lower 3pt\hbox{$\mathchar"218$}}
     \raise 2.0pt\hbox{$\mathchar"13C$}}}
\def\simgt{\mathrel{\spose{\lower 3pt\hbox{$\mathchar"218$}}
     \raise 2.0pt\hbox{$\mathchar"13E$}}}

\def\mic{{$\mu$m}}
\def\h2o{H$_2$O}

\def\teff{$T_{\rm eff}$}

\def\aple{$\mathrel{\hbox{\rlap{\hbox{\lower4pt\hbox{$\sim$}}}\hbox{$<$}}}$}

\def\apge{$\mathrel{\hbox{\rlap{\hbox{\lower4pt\hbox{$\sim$}}}\hbox{$>$}}}$}

\slugcomment{}
\lefthead{Blum {\it et al.}}
\righthead{SAGE-Spec: Extreme AGB Stars}

\begin{document}

\title{ Spitzer SAGE-Spec: Near Infrared Spectroscopy, Dust Shells, and Cool Envelopes in Extreme Large Magellanic Cloud AGB Stars\footnote{Based on observations obtained at the Southern Astrophysical Research (SOAR) telescope, which is a joint project of the Minist\'{e}rio da Ci\^{e}ncia, Tecnologia, e Inova\c{c}\~{a}o (MCTI) da Rep\'{u}blica Federativa do Brasil, the U.S. National Optical Astronomy Observatory (NOAO), the University of North Carolina at Chapel Hill (UNC), and Michigan State University (MSU).}}

\author{R. D. Blum\footnote{NOAO, 950 North Cherry Avenue, Tucson, Arizona, 85719} }

\author{S. Srinivasan, F. Kemper\footnote{Academia Sinica, Institute of Astronomy and Astrophysics, 11F of Astronomy--Mathematics Building, NTU/AS, No. 1, Sec. 4, Roosevelt Rd, Taipei 10617, Taiwan, R.O.C.}}

\author{B. Ling$^{3,}$\footnote{University of Toronto and Academia Sinica Institute for Astronomy and Astrophysics}}

\author{K. Volk\footnote{Space Telescope Science Institute, 3700 San Martin Drive, Baltimore, MD 21218, USA}}

\begin{abstract}

$K-$band spectra are presented for a sample of 39 {\it Spitzer} IRS SAGE-Spec sources in the Large Magellanic Cloud. 
The spectra exhibit characteristics in very good agreement with their positions in the near infrared -- {\it Spitzer} color--magnitude diagrams and their properties as deduced from the {\it Spitzer} IRS spectra. Specifically, the near infrared spectra show strong atomic and molecular features representative of oxygen--rich and carbon--rich asymptotic giant branch stars, respectively. A small subset of stars were chosen from the luminous and red  extreme $``$tip$''$ of the color magnitude diagram. These objects have properties consistent with dusty envelopes but also cool, carbon--rich $``$stellar$''$ cores. Modest amounts of dust mass loss combine with the stellar spectral energy distribution to make these objects appear extreme in their near infrared and mid infrared colors. One object in our sample, HV 915, a known post asymptotic giant branch star of the RV Tau type exhibits CO 2.3 \mic \ band head emission consistent with previous work that demonstrates the object has a circumstellar disk. 

\end{abstract}

{\it Facilities:} \facility{SOAR (), Spitzer()}

\section{Introduction}

\citet{meixner06} presented Surveying the Agents of a Galaxy's Evolution (SAGE), a {\it Spitzer} IRAC and MIPS imaging survey of the Large Magellanic Cloud (LMC). The survey covered the main body of the LMC (7$\times$7 degrees$^2$) in all the IRAC and MIPS bands. The resolved stellar populations in the LMC were particularly well covered on the giant and asymptotic giant branches (AGB) as well as on the sequences of more luminous red supergiants (RSG) \citep{blum06}. \citet{kemper10} selected a sample of $\sim$ 200 objects from the SAGE color--magnitude diagrams (CMD) to observe with the {\it Spitzer} IRS (see below); this IRS data set is known as the SAGE--Spec survey. See also \citet{zijlstra06} and \citet{sloan08} for {\it Spitzer} IRS observations of evolved stars in the LMC. In this paper, we present near infrared $K-$band spectra of the brighter IRS sources from the \citet{kemper10} sample. The IRS spectra have been classified with respect to their basic properties according to their mid and far--infrared spectral characteristics  by \citet{wood11}.

One of the primary goals of SAGE and SAGE--Spec is to quantify the mass loss return of evolved stars into the LMC interstellar medium. To this end, we have embarked on a program to study individual objects representing each class of mass losing star in the CMD to better understand the mass loss contributions as a function of position on the CMD (i.e. by evolutionary state, chemical composition, and mass). Knowledge of the type of $``$stellar core$''$ embedded in the dusty envelope for any given source is an important model input and can help elucidate the nature of the object as well. Both \citet{srini10} and \citet{sargent10} have modeled individual sources from the SAGE/SAGE--Spec sample. In these cases, photometry was used to constrain the stellar models at optical and near infrared wavelengths. Adding near infrared spectra can further constrain the stellar core properties (for example, leading to a more accurate $T_{eff}$). 

The SAGE--Spec sample included 197 sources selected from the SAGE IRAC and MIPS photometry \citep{kemper10}. The sample was selected to cover the range of populations (young and old) expected based on the known star--formation history of the LMC and cover each area of the CMD. The present near infrared program concentrated on the brighter objects in the sample and those concentrated on the AGB, RSG, and extreme AGB (XAGB) as defined by \citet{blum06}. These three populations are easily seen in Figure~\ref{cmd} and the sub--sample of objects observed is over plotted in the diagram. The RSG are the sequence with the brightest objects observed in this sample (e.g. object 19). The AGB stars include both oxygen (O--) rich and carbon (C--) rich stars above the red giant branch (the  dense sequence of stars at $J - $[8.0] $\approx$ 1.0 and reaching [8.0] $=$ 12). The XAGB starts near object 69 in this diagram and extends to redder colors (to object 144 and beyond). Galaxies inhabit the broad clump of fainter objects near $J-$[8.0] $\approx$ 5.0. The obvious sequences to the blue of the red giants and RSG are warmer evolved stars, hot stars, and foreground objects; see \citet{blum06} for details.

\section{Observations}

The brighter stars at 2 \mic \ ($K$ band) were chosen from the SAGE--Spec sample to be observed at the SOAR 4.1--m telescope on Cerro Pach\'on in Chile. Practically speaking this meant sources with $K \le$ 12 mag. A sample of 39 stars was observed over a five night period in November, 2007.
Spectra were obtained with the Ohio State InfraRed Imaging Spectrograph, OSIRIS. OSIRIS is described in detail by \citet{depoy93}. The instrument was configured in the $f$/7 camera mode, which delivers 0.14$''$ pixel$^{-1}$, an 0.45$''$ (3 pixel) slit and a spectral resolution of $R=$ $\lambda/\Delta\lambda$ $=$ 3000.

Data were taken on the nights of 16--20 November 2007 (local time). The weather was good, mostly clear for the entire five night run as indicated by the GOES satellite imaging archive maintained at CTIO\footnote{http://www.ctio.noao.edu/sitetests/GOES/}. The seeing was also good as measured on acquisition images. 16 November was outstanding with acquisition images recording delivered image quality (DIQ) under 0.5$''$ most of the night. The other nights had DIQ near 0.5$''$ in acquisition images most of the time. Excursions to 0.7$''$ were observed as was one acquisition sequence as good as 0.31$''$. The objects were identified unambiguously in 2MASS images using the SAGE coordinates. In each case a bright $K-$band source was easily identified as the target in acquisition images which we compared to the 2MASS image. SAGE IRAC sources have typical positional differences from 2MASS positions of \aple \ 0.3$''$ \citep{meixner06}.

Each source was observed in the same way. Objects were acquired in imaging mode with OSIRIS and then placed on the center of the 72$''$ $\times$ 0.42$''$ slit. The telescope was then moved to dither the source along the slit obtaining six samples separated by 6$''$. Observations of hot stars (A--type) were made throughout each night every few hours and at airmass within 0.1 of each object to be used for correcting Telluric features in the object spectra.

Flat field images were obtained using the calibration lamp mounted on the same Instrument Support Box (ISB) that the instrument is mounted on. Flats were taken after each object sequence in the hope that this would more effectively remove fringing in the spectrum due to the instrument. However, this was not the case and in the end flats were averaged throughout the night. 

\section{Data Reduction}

All basic data reduction steps were completed with IRAF\footnote{IRAF is distributed by the National Optical Astronomy Observatory, which is operated by the Association of Universities for Research in Astronomy (AURA) under cooperative agreement with the National Science Foundation.}. IRAF wrapper scripts from the CTIO InfraRed Reduction 
(CIRRED\footnote{http://www.ctio.noao.edu/instruments/ir$\_$instruments/datared.html}) package were used to execute 
the actual data reduction steps.

The object images were flattened with the dome flats, and bad pixels corrected by interpolating across them in each dimension. The bad pixel mask was made using illuminated flat field images and dark images. 

Sky images were made by median combining the on--source images themselves.  Once corrected by the flat field and bad pixel mask, each object image was sky subtracted using the median sky image from that object sequence. One dimensional spectra were extracted from these images, and the six samples combined to form an object spectrum. The object spectra were calibrated in wavelength using the quadratic solution to emission lines from an Argon lamp in the ISB calibration source. The RMS error in the known positions of Ar lines was less than 0.1 pixels (1 pixel $=$ 3.7 \AA). All the spectra were arbitrarily shifted to have a CO 2-1 rotational vibrational band head position of 2.2935 \mic.

One dimensional spectra were extracted from the two dimensional images using IRAF APEXTRACT. The average spectral flux differences between individual dithers (typically 10 $-$ 20 $\%$ but with occasional larger excursions) are dominated by slit positioning (due to tracking, flexure, and slit position angle orientation error) and variable cloud (in some cases). Each spectrum from individual dithers was normalized before median combining to enhance bad pixel rejection.

Each object spectrum was then divided by a hot star telluric $''$standard$''$ star spectrum (HD or Hipparcos A--type stars) that had been reduced in just the same way. Small shifts in wavelength were made to align the hot star telluric features with those in the object spectrum (by cross correlation). The hot star chosen was always one that had been observed at an airmass within 0.1 of the object, and usually within 1 to 2 hours clock time. 

As noted by \citet{dewitt13}, there is an unknown fringe or other instrumental signature in the OSIRIS spectra taken at SOAR (see Figure~\ref{map}). The pattern manifests itself as three main features approximately 60 pixels in width with amplitude 10$\%$ to 20$\%$ of the continuum. There are secondary features associated with each primary feature that have $\sim$ half the primary amplitude. Fortunately, these features are straightforward to remove from our spectra at a level which is sufficient for our science goals. The features appear at times, but not other times and affect both the objects and standards in the same way. 

The features are removed as the last step of the reduction process. First, for each night, we built two feature or fringe maps. These were obtained by dividing hot star spectra of the same star taken at different times and then smoothing the resulting ratio by a $\sim$ 100 pixel moving average. The only structure left in such a ratio is the feature strength as a function of wavelength. Since we don't know if any particular object will have $''$emission$''$ or $''$absorption$''$ features after dividing by the telluric standard, we construct a map of both types. The telluric corrected object spectrum was reviewed and then the appropriate map chosen to use to correct any of the fringe--like features. The correction was made by using the IRAF task TELLURiC which allows the user to interactively shift a standard spectrum (in this case the feature map) relative to the object and also vary the strength of the features. Each object was corrected in this way and the spectra shifted and feature strength adjusted by eye until the best final spectrum was obtained (an RMS difference is given by the task TELLURIC which was also used to guide the choice of final correction). While the source of the features remains unclear, the correction is relatively straight forward and relies entirely on taking multiple spectra of the same object. An example of the features, uncorrected and corrected spectra is shown in Figure~\ref{map}. 

\section{Results}

The object properties are listed in Table~\ref{point}. The objects are listed in order of increasing $J-$[8.0] color. The first column gives the object number as defined in our original observing list. The sources were independently numbered in the {\it Spitzer} IRS proposal. Each spectrum has an associated ID for the {\it Spitzer} IRS observations \citep{wood11} as well, and this SSID is noted in the table. A $``$type$''$ is assigned in the table based upon  three distinct observations: the location in the CMD, the (present) near infrared spectrum, and the point source classification of \citet{wood11} for the associated IRS SSID. 

The great majority of the sources presented herein are known variable stars; see the descriptions presented for each object by \citet{wood11} based on the variable classifications of \citet{alcock98, groen04, ita04, fraser05, fraser08, sosz09} and \citet{vijh09}. Riebel and collaborators (private communication) have made a study a the photometric variability of SAGE sources in the IRAC bands. They find large amplitude variations (0.3 -- 1 mag) for objects 3, 84, 87, 88, 94, and 144 in our sample in [3.6] and [4.5].  

\begin{figure}
\plotone{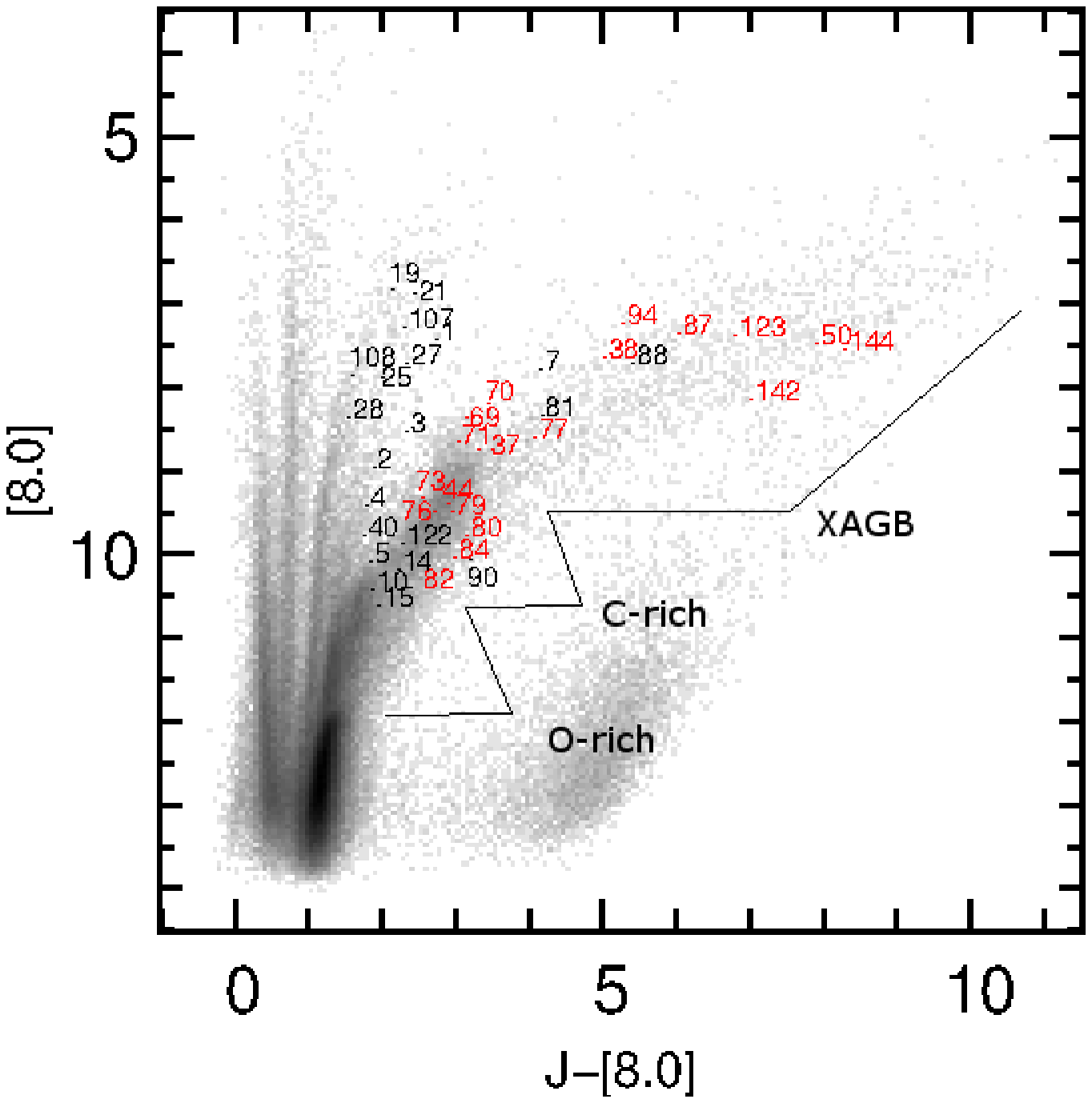} 
\caption{$J-$[8.0] $vs.$ [8.0] color--magnitude diagram (CMD) from the SAGE Survey \citep{meixner06,blum06}. The CMD uses the 2MASS near infrared $J$ magnitude \citep{skrutskie06}. The SAGE--Spec sub--sample of red supergiants, O--rich AGB stars, C--rich AGB stars and XAGB stars observed here are labeled as {\it black} for O--rich and {\it red} for C--rich stars in the online version of the paper. The various sequences in this diagram are discussed in \citet{blum06} and the text. The density of objects has a maximum of 1062 sources per 0.06 mag $\times$ 0.06 mag bin. See Figure~\ref{blow} for a blow up of the asymptotic giant branch (AGB), red supergiant sequence, and extreme AGB with the same labels.
\label{cmd}}
\end{figure}

\begin{figure}
\plotone{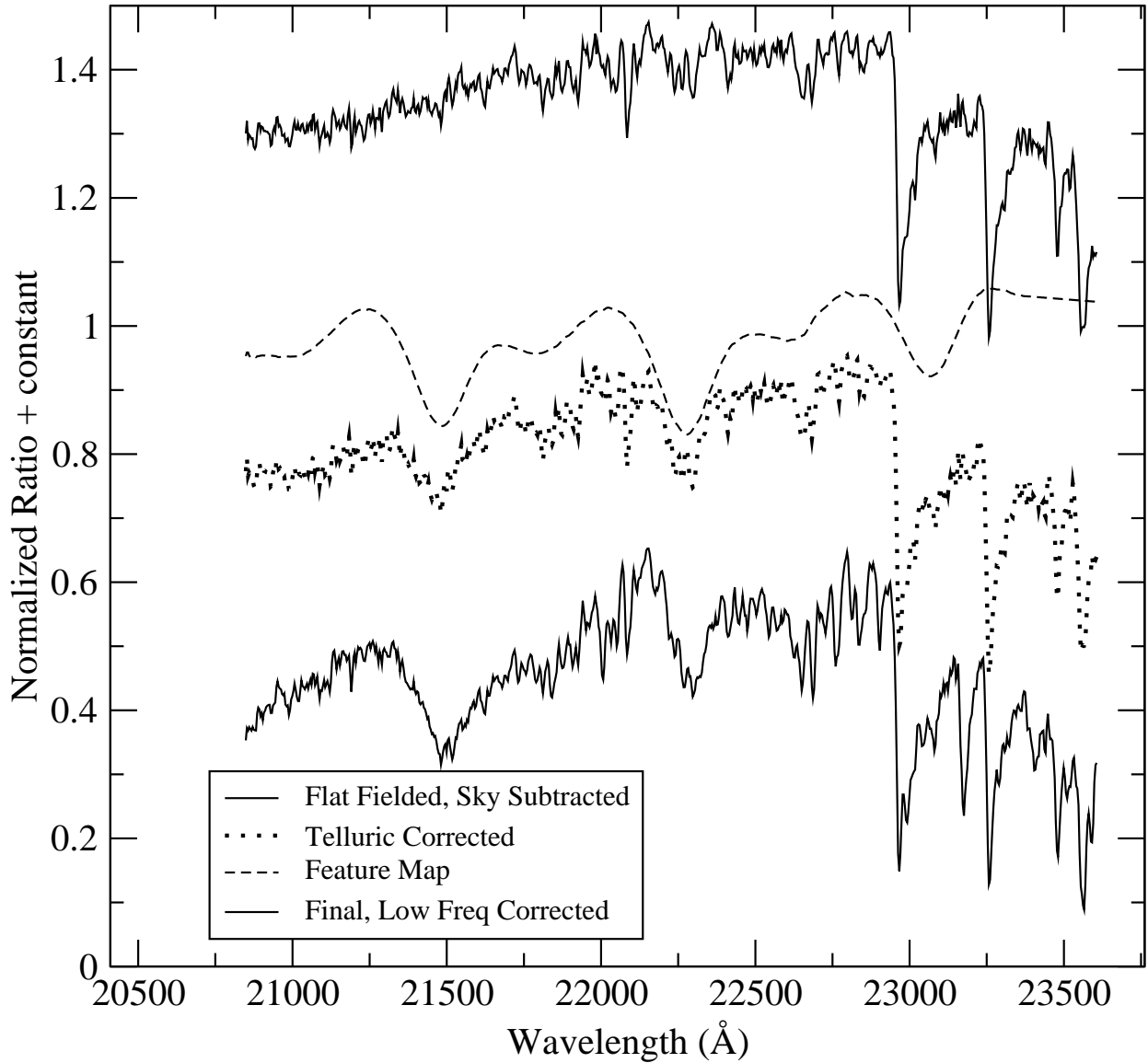} 
\caption{Example object spectrum (obj4) with fringe--like features: {\it solid} line, object spectrum; {\it dotted} line, object corrected by telluric standard; {\it dashed} line, fringe--like feature map; {\it solid} line, final corrected spectrum. The feature map was constructed by dividing a spectrum of a hot star with a spectrum of itself obtained on the same night. See text.
\label{map}}
\end{figure}

\begin{figure}
\plotone{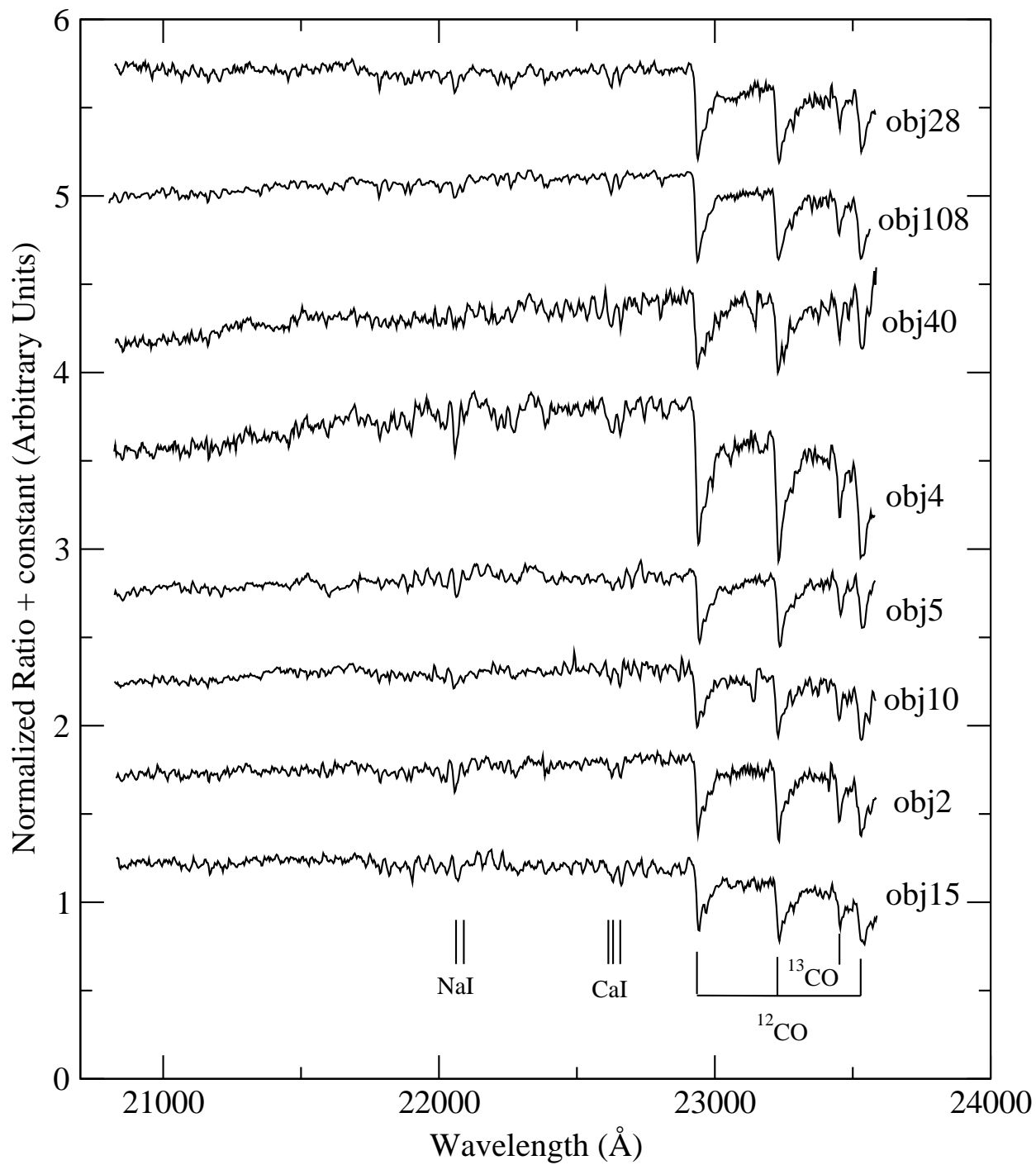}
\caption{Objects classified as O--rich based upon their near infrared spectra presented herein. Objects are ordered on $J-$[8.0] from blue to red (top to bottom, respectively). The spectra exhibit normal M--type or O--rich type features such as Na, Ca, and CO absorption;  see text and Table~\ref{point}. The difference between normal O--rich and normal C--rich stars is easily seen by comparing this Figure and Figure~\ref{orich2} to Figure~\ref{crich}, all plotted on the same scale. 
\label{orich1}}
\end{figure}

\begin{figure}
\plotone{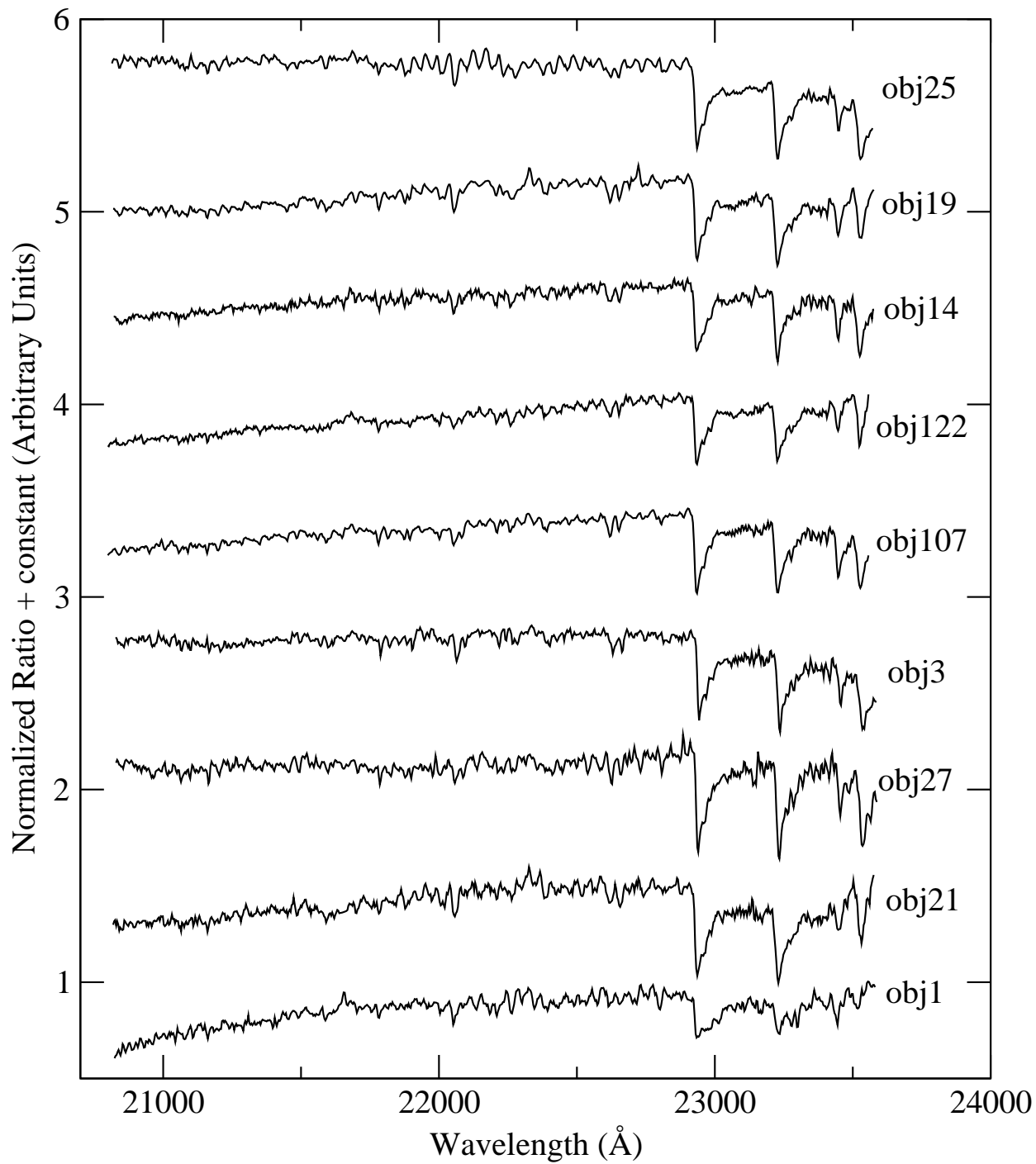}
\caption{Same as Figure~\ref{orich1} but for the remainder of the sample of O--rich stars. These are ordered by $J-$[8.0] from blue to red (top to bottom, respectively) and all with $J-$[8.0] more red than the objects in Figure~\ref{orich1}; see Table~\ref{point}.
\label{orich2}}
\end{figure}

\begin{figure}
\plotone{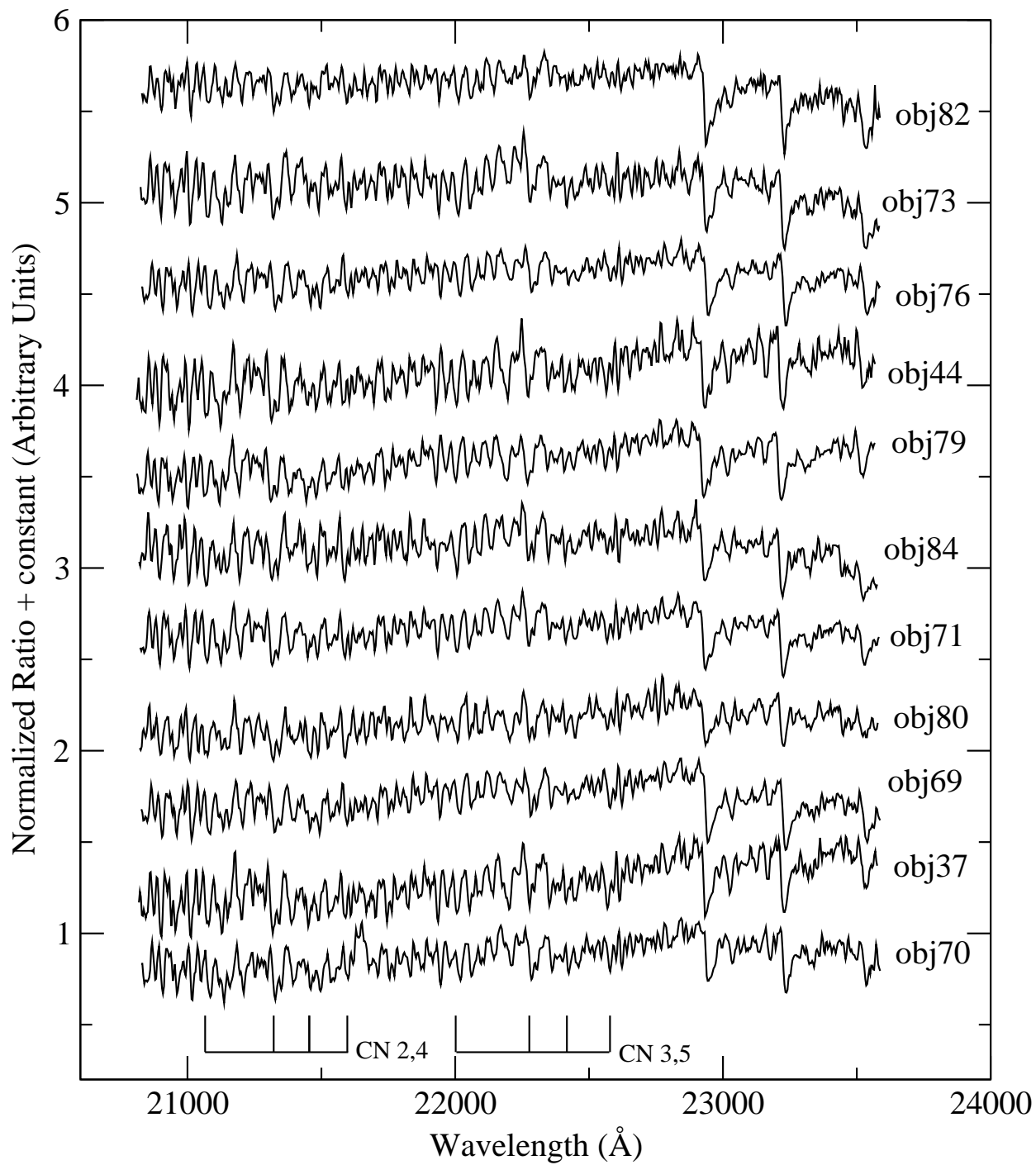} 
\caption{Same as Figure~\ref{orich1} but for stars classified herein as C--rich. Positions of the $\nu =$ 2--4 and 3--5 band heads of CN are show \citep{ts70,wh96}.
\label{crich}}
\end{figure}

\begin{figure}
\plotone{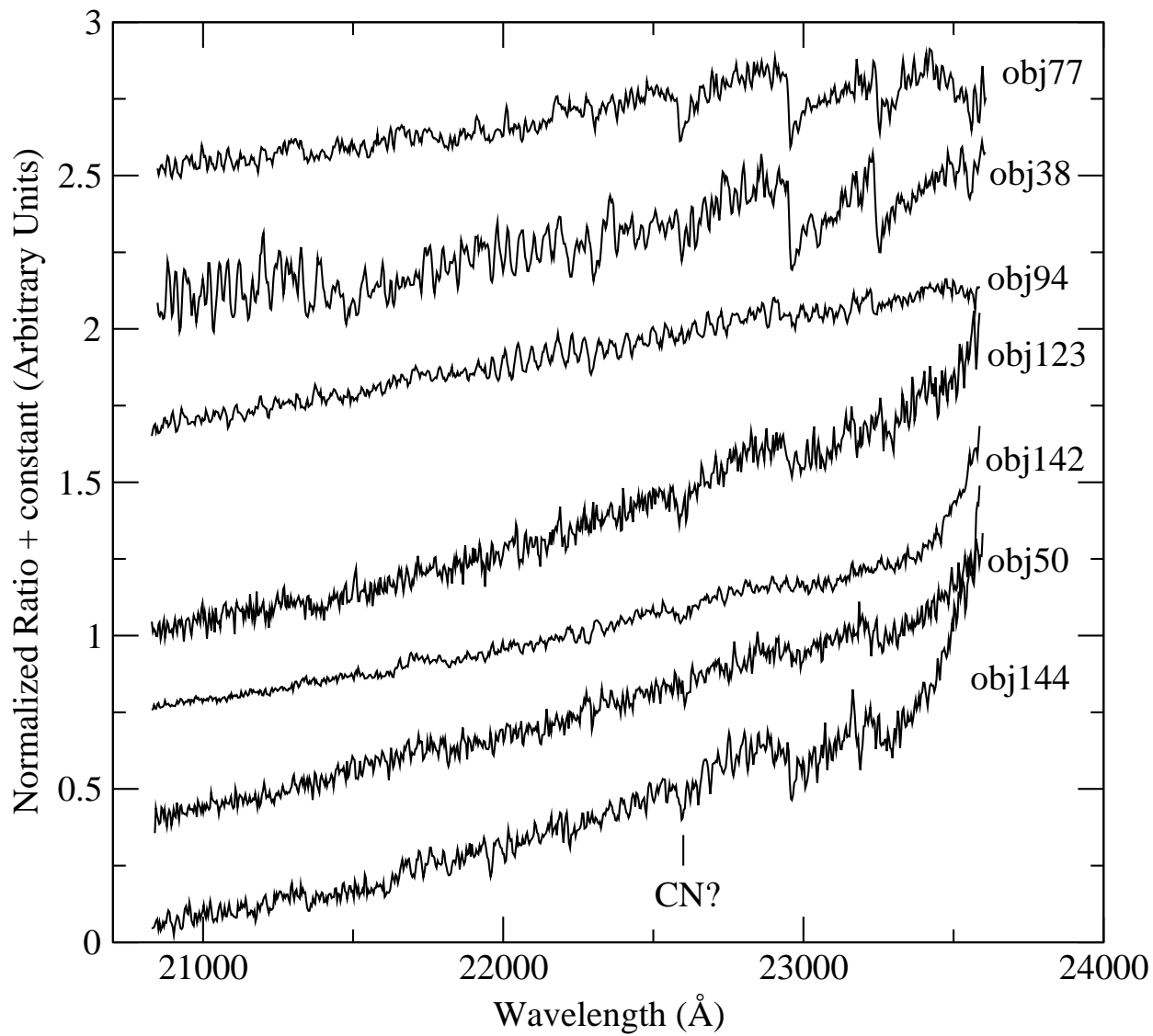} 
\caption{Extreme asymptotic giant branch (XAGB) stars. Such stars have $J-$[8.0] $>$ 3.1; see text. A feature, possibly due to CN, is seen in most of the objects near 2.26 \mic. The feature may arise in the extended envelopes of the mass losing stars; see text. Object 94 was selected for observing with the {\it Spitzer} IRS, but the observation resulted in a different object being observed by {\it Spitzer}. Thus no IRS counterpart exists. Object 94 is essentially featureless. The strong fringing in the central part of the spectrum is instrumental; see text.
\label{x}}
\end{figure}

\begin{figure}
\plotone{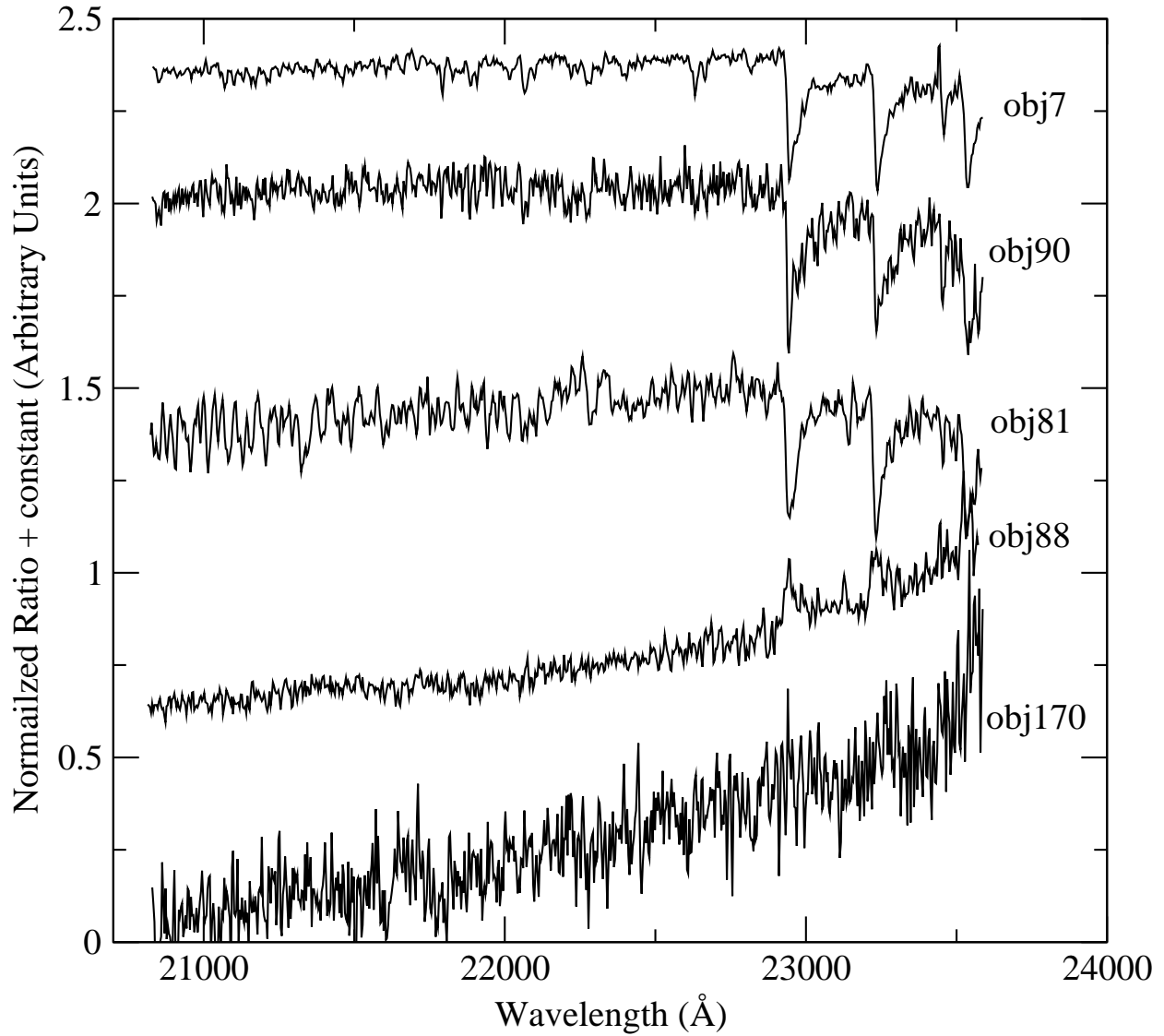} 
\caption{Objects whose near infrared spectra are not typical of their position in the $J-$[8.0] $vs.$ [8.0] CMD or are not typical asymptotic giant branch (AGB) stars. Object 7 is a normal O--rich or M--type star that sits above the XAGB sequence in Figure~\ref{cmd}; see text. Objects 81 and 88 are O--rich post--AGB stars and 88 exhibits CO 2.29 \mic \ emission. Object 170 is an external (to the Large Magellanic Cloud) galaxy according to \citet{wood11}, consistent with its position in the CMD \citep{blum06}.
\label{other}}
\end{figure}

The object spectra are shown in Figures~\ref{orich1}, \ref{orich2}, \ref{crich}, \ref{x}, and \ref{other}. In the following, we discuss the near infrared features in these spectra that clearly identify their O--rich, C--rich, and ``extreme'' character.
Inspection and comparison of the individual spectra with the labels in Figure~\ref{cmd} for the $J-$[8.0] CMD shows the spectral morphology indicated is well matched to the position of the objects in the CMD. For example, C--rich stars do not appear on the lower part of the AGB in Figure~1 or on the RSG sequence. O--rich objects are generally confined to the lower part of the AGB, and the RSG sequence appears to be exclusively populated by O--rich stars. Some O--rich objects do fall on or near the XAGB, and we discuss these further below.

The objects shown in Figures~\ref{orich1} and \ref{orich2} have the distinct features of normal M--type stars that are typical of stars on the RSG locus and O--rich AGB locus \citep{niko00, cioni06, blum06}. The features include strong CO absorption due to the 2--0 rotational--vibrational band head at 2.29\ \mic, as well as atomic features such as \ion{Na}{1} and \ion{Ca}{1} at 2.21 \mic \ and 2.26 \mic, respectively \citep{kh86}. All these objects are independently identified as O--rich by \citet{wood11} from their {\it Spitzer} IRS spectra.  

The C--rich stars (Figure~\ref{crich}) exhibit strong, unresolved, molecular absorption that results in so much distributed opacity the continuum appears $``$noisy$''$. But this is not the case (as is well known), the spectra are normal for these types; see, for example, \citet{fh72}, or more recently the compilation of \citet{tanaka07}.  The C--rich objects do still show strong CO absorption at 2.29 \mic, but atomic features typical in the O--rich stars are not identifiable. Even a cursory comparison of Figures ~\ref{orich1}, \ref{orich2}, and \ref{crich} shows that the stars are easily distinguished by their near infrared spectra. All these objects are independently identified as C--rich by \citet{wood11} from their {\it Spitzer} IRS spectra.  

The stars depicted in Figure~\ref{x} are so called $``$extreme AGB (XAGB)$''$ stars as defined in \citet{blum06}. These objects populate the plume of AGB stars that run to very red ($``$extreme$''$) colors in the $J-$[8.0] $vs.$ [8.0] CMD (Figure~\ref{cmd}). As defined by \citet{blum06} and based upon near infrared colors \citep{niko00, cioni06}, the XAGB stars are those with $J-$[8.0] $>$ 3.1 mag. The sequence of spectra goes toward redder $J-$[8.0] from top to bottom. The bluest XAGB stars appear like normal C--rich objects. Those toward the red end of the sequence, however, become nearly featureless. There is still apparent CO 2.29 \mic \ absorption, though it is weaker as is the presumed normal distributed absorption seen in Figure~\ref{crich}. \citet{fh72} studied Galactic C--stars with a range of colors and concluded veiling of the stellar photosphere by dust emission arising in the shell would explain the anti correlation of CO absorption strength and infrared color. None of the \citet{fh72} stars appears as $``$veiled$''$ as the reddest XAGB stars presented here or with as weak absorption, but the veiling mechanism is probably correct in the LMC case as well (see \S5).

The XAGB objects all show a strong feature not present in the normal C--rich stars, absorption at 2.26 \mic. This may be due to CN \citep{wh96}.  \citet{fh72} also identify numerous absorption features due to CN (some near 2.26 \mic) as well as features due to C$_2$H$_2$, and HCN, though no feature as strong as shown in Figure~\ref{x} is present at 2.26 \mic \ in their spectra. \citet{wood11} classify the XAGB stars in Figure~\ref{x} as C--rich from their mid--infrared IRS spectra.

The remaining few stars are shown in Figure~\ref{other}. These objects, discussed below, have spectra that are not necessarily expected for their location in the CMD. It is only with additional information \citep[the {\it Spitzer} IRS point source characterization]{wood11} that their nature is discovered. Object 170 is a galaxy according to \citet{wood11}. Its spectral energy distribution (SED) places it in the galaxy region of the CMD \citep{blum06}. The remaining four objects turn out to be O--rich, though only object 7 is obviously so from the near infrared spectrum. Objects 7, 81, 88, and 90 are classified as post--AGB (PAGB) stars \citep{wood11}. Object 88 shows CO emission in the near infrared, while objects 81 and 90 show CO 2.3 \mic \ absorption. 

\section{Discussion}

\citet{blum06} categorized most of the evolved stars in the LMC {\it Spitzer} and 2MASS CMDs based on colors and magnitudes and confirmed the basic characterization of objects in Figure~\ref{cmd}  through association of position with what was known from spectroscopic follow--up in the literature. More recently, \citet{boyer11} did a similar classification for the AGB stars in the Small Magellanic Cloud which results in an an analogous set of C--rich and O--rich AGB stars. \citet{boyer11} adopted the same color cut to define XAGB stars in the $J-$[8.0] $vs.$ [8.0] CMD as did \citet{blum06}. As seen in Figure~\ref{cmd} \citep[adopted from][]{blum06}, the XAGB clearly appears to be an extension of the AGB which itself had two distinct O-- and C--rich components. The RSGs were classified as a luminous branch of more massive O--rich (i.e. K or M--type) stars and between the AGB and RSG sequence were bright or luminous O--rich AGB stars. But at colors with $J-$[8.0] \apge 3, little was known, and it was not clear what the nature of any particular XAGB star should be. The main objective of this work is to use near infrared spectra and the {\it Spitzer} IRS spectral characteristics to further explore the AGB with emphasis on the XAGB.

Figure~\ref{blow} shows a blowup of the AGB, RSG, bright O--rich, XAGB region of the $J-$[8.0] $vs.$ [8.0] CMD. The spectra presented in this work were selected to fall neatly on the sequences in this diagram. The O--rich stars (Figures~\ref{orich1} and \ref{orich2}) are distributed on the RSG sequence, the bright O--rich AGB sequence (between the RSGs and the normal C--star sequence), and at the tip of the main O--rich AGB above the red giant branch.

The normal C--rich stars (Figure~\ref{crich}) fall on the main C--rich sequence. XAGB objects 77 and 38 (Figure~\ref{x}) are clearly normal C--rich stars (compare Figure~\ref{crich}). The IRS spectra are consistent with this classification. Of the remaining objects with colors that place them on or near the XAGB, most have rising continua to the red and weak features (objects 94, 87, 123, 142, 50, 144). One object, number 87 was not detected (we will discuss it briefly below). There are apparently three O--rich objects in the XAGB, and these are discussed in \S~5.1. 

\subsection{Extreme AGB Stars}

As noted above, several of the XAGB objects have near infrared spectra of normal C--stars. Upon closer examination, most of the XAGB stars with weak features have absorption band heads due to CO. Only object 94 and 142 do not (or the feature is very weak). All of the XAGB stars have evidence of an absorption feature at 2.26 \mic, except object 94. Above we speculated this absorption might be due to CN. 

The objects in the XAGB also have the reddest colors in our sample. Their near infrared spectra rise steeply toward longer wavelengths. The nature of the continuum combined with the absorption due to CO and the fact that several are normal C--stars suggests the XAGB is largely populated with C--stars with varying amounts of dust emission. This emission causes the spectra to be rising in the near infrared and veils, or dilutes the otherwise normal C--star spectrum. \citet{fh72} used the anti--correlation of CO 2.29 \mic \ absorption strength and infrared color to argue this veiling mechanism was responsible for the systematic change in spectral morphology of a sample of Galactic C--stars.  

\begin{figure}
\plotone{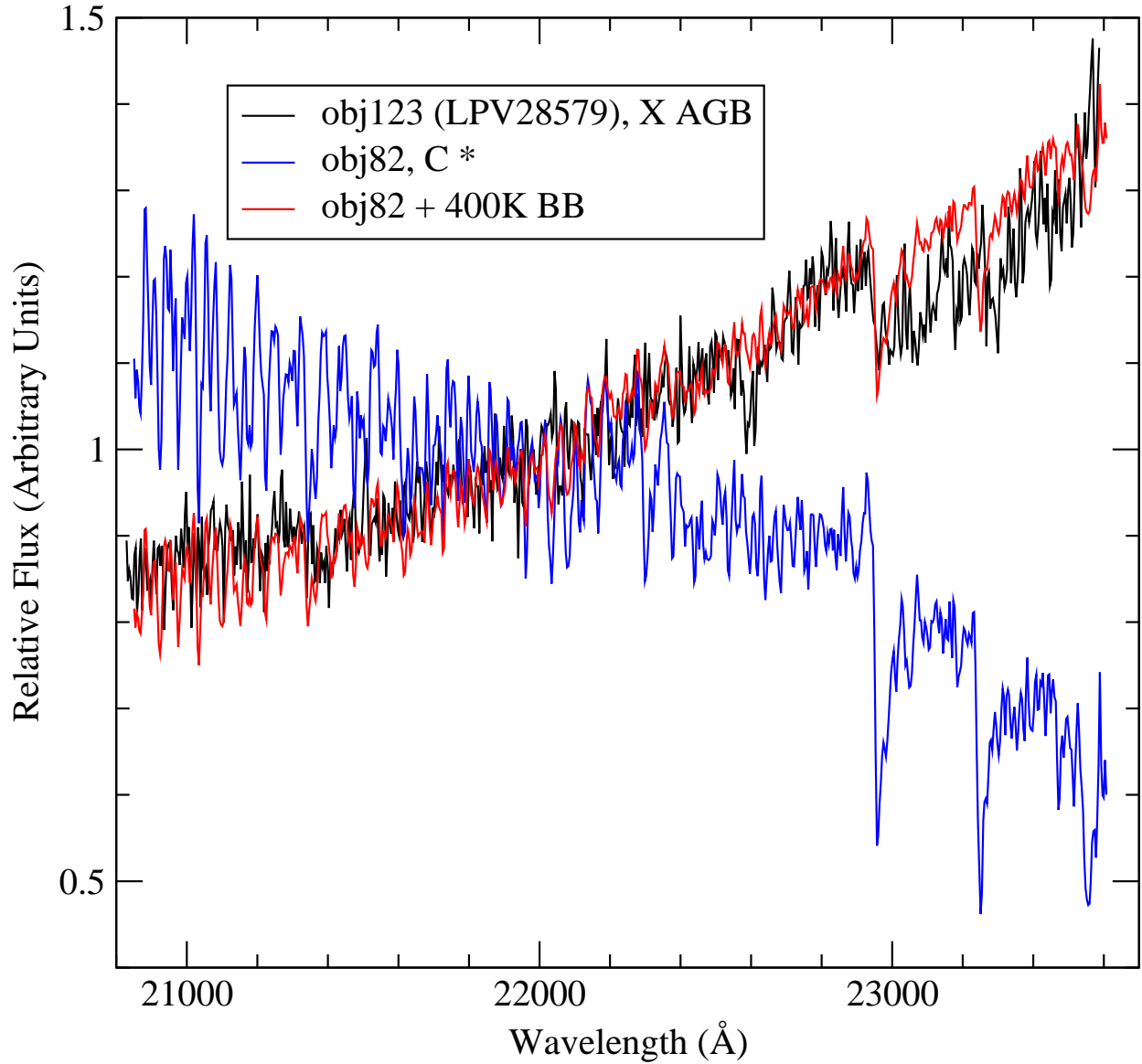}
\caption{One of the four most extreme AGB stars in our sample, object 123 ({\it black} line). The continuum rises steeply to the red. Molecular features appear to be $``$suppressed$''$ or veiled relative to more normal C--rich stars (see Figure \ref{crich}). A simple combination of a normal C--rich star (object 82, {\it blue} line) and a black body of 400~K in equal proportion ({\it red} line) match the extreme AGB star very well. The CO feature is veiled and the continuum slope is similar. There is a strong feature at 2.26 \mic \ that does not appear in the normal C--rich star. The line may be due to CN; see text. The simple model does not include reddening, so the blackbody is expected to be cooler than it otherwise would be (see text and Figure~\ref{2dustb}. This figure appears in color in the online version of the paper.
\label{xagbbb}}
\end{figure}

\begin{figure}
\plotone{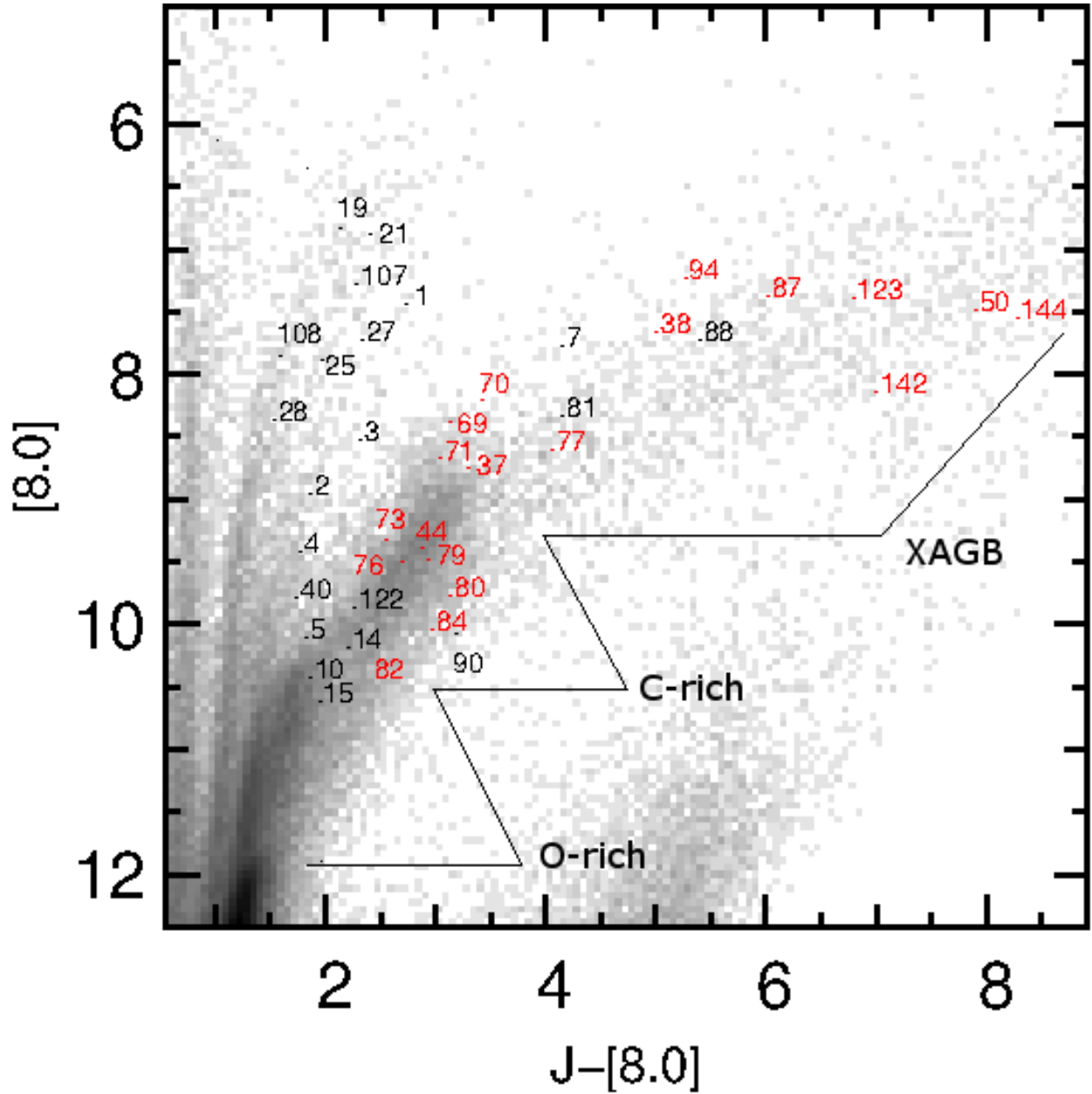} 
\caption{Blow up of the extreme asymptotic giant branch (XAGB) for the $J-$[8.0] $vs.$ [8.0] color--magnitude diagram shown in Figure~\ref{cmd}. Object labels are the same as Figure~\ref{cmd}. The O--rich, C--rich, and XAGB sequences are indicated approximately in the Figure; see text. This figure appears in color in the online version of the paper.
\label{blow}}
\end{figure}

To further test this idea, we took one of our normal C--stars and combined its near infrared spectrum with an equal contribution from a cool black body as a simple representation of emission from a dust shell (trial and error yielded a good match with a 400K blackbody). The result is shown in Figure~\ref{xagbbb}. The combined spectrum is a remarkably good match to one of the XAGB stars, object 123. The unknown absorption feature is obviously not reproduced since it is not present in the original spectrum of the normal C--star, object 82. However, the veiling of the myriad absorption lines in the C--star including CO at band heads at 2.3 \mic \ and longer is well reproduced as is the overall shape of the continuum. The apparently good fit is fortuitous since there is also reddening in the 
circumstellar environment which will cause the near infrared spectrum to be more red. The 400K BB overcompensates (i.e. it is cooler than it should be in the absence of reddening) for the required slope, and so we might expect the real temperature to be higher (this is indeed the case as shown by detailed modeling; see below).

It is tempting to assume the reddest objects in the present sample must have the strongest mass--loss. However, the {\it Spitzer} IRS spectra suggest weak dust emission features for the XAGB objects observed at SOAR as a whole. The brighter  O--rich stars in the present sample show more prominent emission features in the mid--infrared due to dust. The SOAR sample is limited by objects whose $K-$band magnitudes are brighter than 12, so the present sample, while very red in $J-$[8.0], does not contain the reddest XAGB stars \citep[see for example][]{gruendl08,riebel12}.

In order to verify that the picture of the XAGB outlined above is correct, we ran a series of models to test whether modest amounts of dust could reproduce both the near infrared spectra of the XAGB stars (as suggested by Figure~\ref{xagbbb}) as well as the mid--infrared {\it Spitzer} IRS spectra. Our models were created using the dust code 2Dust by \citet{ueta03}. We represent the central star in these models with the \citet{aringer09} carbon--star models of LMC metallicity. 

%-------------------------------
The present analysis expands on the work of \citet{srini10} who did a detailed model of object 123 (but without specific knowledge of the near infrared spectrum). In the present work, we test the nature of the XAGB by modeling one of the most "extreme" cases, object 142 (Figure~\ref{x}). Object 142 was defined as a variable star by \citet{vijh09} based on its SAGE epoch 1 and 2 IRAC photometry which varied significantly compared to photometric uncertainties. The IRAC epoch 1 data were obtained in 2005 July and epoch 2 followed about three months later. The MIPS data for each epoch were obtained in the week following the corresponding IRAC data. In addition to the SAGE photometry, near infrared photometry is available from 2MASS \citep{skrutskie06} and the InfraRed Survey Facility \citep[IRSF]{kato07}. Visible photometry (V and I) is available  from \citet{zar04}. None of these observations was obtained simultaneously with another.
 
The dust properties were kept fixed, and were the same as in the GRAMS carbon--star grid (Srinivasan et al. 2011); in particular, the dust consisted of a mixture of amorphous carbon and 10$\%$ SiC by mass. Using a range of stellar cores with \teff \ between 3000 K and 4000 K, we constructed dust shell/envelope models with varying amounts of dust (controlled by the strength of the SiC feature at 11.3 \mic) and a few different values for the inner radius of the dust shell. 

\begin{figure}
\plotone{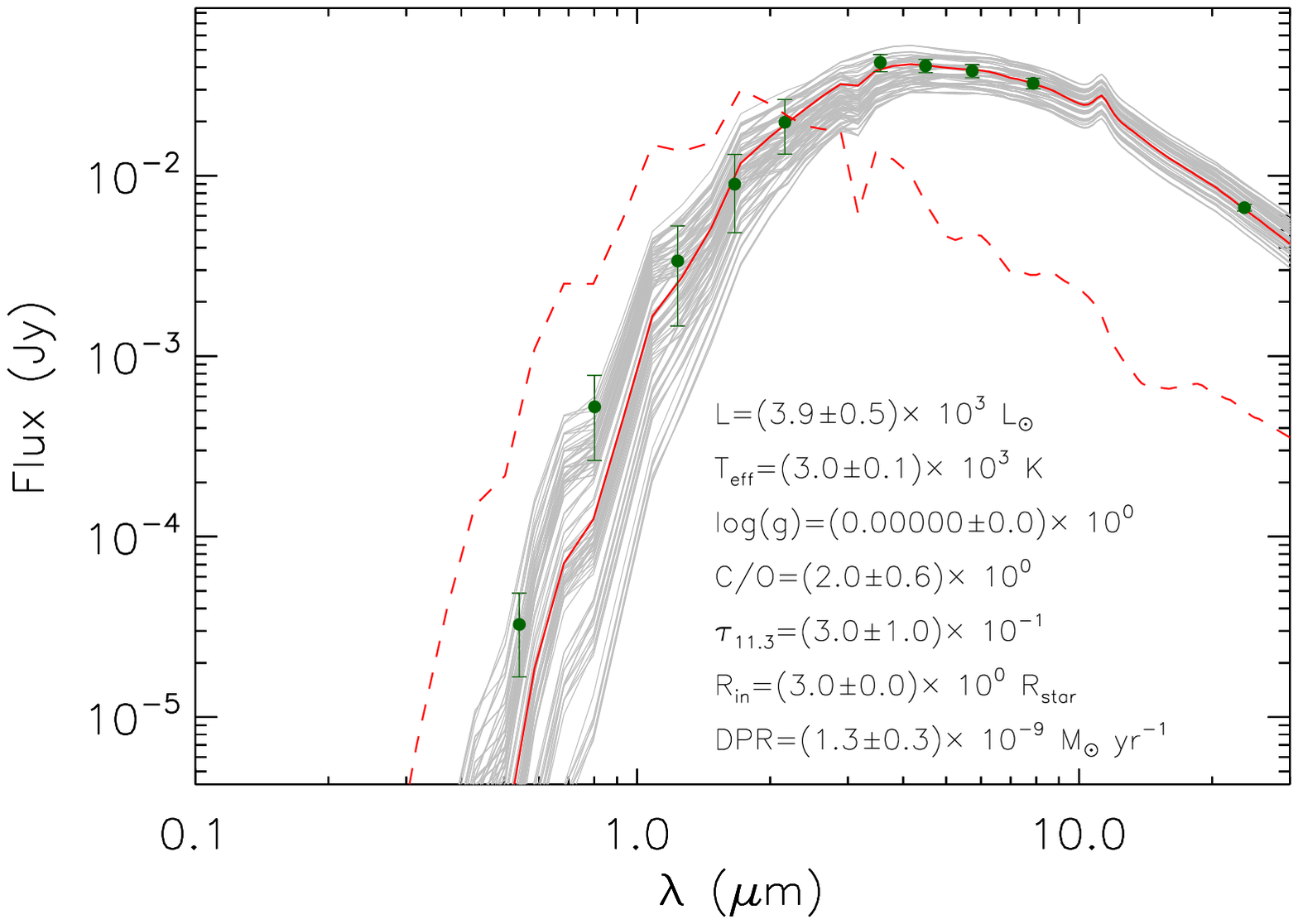} 
\caption{2Dust models \citep{ueta03} compared to the observed mean photometry for object 142.  The data points show the mean MIPS, IRAC, near infrared (2MASS and IRSF), and visible photometry \citep{zar04}. The {\it dashed red} line is the input model photosphere, and the {\it solid red} line is the best fit to the mean photometry. Best fits to the mean photometry have optical depth in the SiC 11.3 \mic \ line of about 0.3. The range of optical depths represented by the grey lines in the figure is 0.2 -- 0.4. These models have core photospheres with \teff \ $=$ 3000K; see text. 
\label{2dusta}}
\end{figure}

Models were constructed for two distinct cases for object 142. In the first case, we compared models only to the mean photometry. The results are shown in Figure~\ref{2dusta}. 
We find a range of models that are consistent with the photometry taking into account variability for the multiple epochs of the data. Models with \teff \ between 3000K and 4000K and with optical depth in the SiC 11.3 \mic \ line of 0.2 to 0.4 were constructed and compared to the data. Best fits to the mean photometry were obtained with \teff \ $\approx$ 3000K and optical depth of 0.3. W found that the inner radius was reasonably constrained by the mid-IR photometry alone \citep[as was the case for][]{srini10}. The best match occurs for an inner radius equal to three times the stellar radius (corresponding to inner radius dust temperatures near 1000 K). These models are normalized to the IRAC [8.0] mag flux.

The second case compared the model results to photometry, the SAGE--Spec IRS spectrum, and the present near infrared spectrum. The goal was to see if a model could reproduce the dusty IRS spectrum including the SiC 11.3 \mic \ feature, the veiled near infrared spectrum and the photometry.
We modified the near--infrared part of the input carbon--star photospheric  models used with 2Dust by substituting the scaled  $K-$band spectrum of  object 82 for the model flux in this region. Object 82 is a normal C-star that is assumed to have negligible reddening from its circumstellar envelope or the interstellar medium of the LMC (at least in the near infrared). By adding in the normal C--star spectrum to the core star we can see how the emergent spectrum compares to the observed XAGB spectrum of 
object 142.

As a consistency check, we ran a set models to compare to the photometry and SAGE--Spec spectrum of object 82. The object 82 data are best fit by models with similar \teff \ and luminosity as we found for the fits to the mean photometry for object 142 (3200K $\pm$ 100K $vs.$ 3000K $\pm$ 100K and 
51000 $\pm$ 600 L$_{\odot}$ $vs.$ 39000 $\pm$ 500 L$_{\odot}$ for object 82 and object 142, respectively).

Initial comparisons of the models to the data showed that no models could fit the IRS data, the MIPS photometry and simultaneously the mean near infrared and visible photometry {\it and produce a veiled near infrared spectrum} at the same time. Completely veiling the near infrared spectral features required more dust than the models produced while still reproducing the (brighter) short wavelength fluxes. However, models could be obtained which compared well to the IRS spectrum, a sub--set of the near infrared, IRAC, and MIPS photometry and which simultaneously produced a suitably veiled near infrared spectrum.

\begin{figure}
\plotone{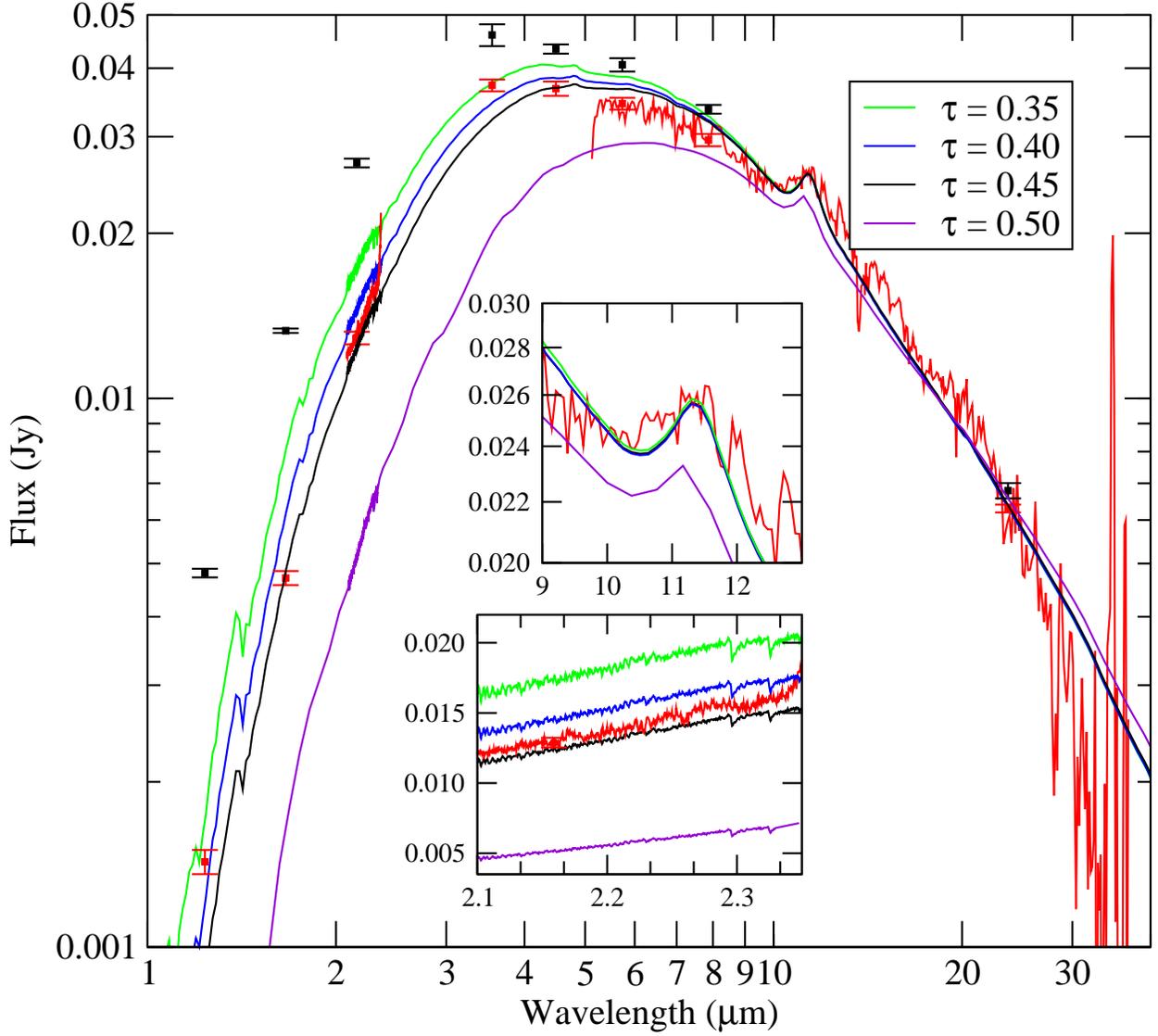} 
\caption{2Dust models \citep{ueta03} and observed fluxes for object 142. {\it Red} data points and spectra are compared to the models. {\it Black} data points are for reference.
The models were constructed with an input photosphere that incorporates the near--infrared spectrum of the normal C--star object 82 (see the Figure {\it lower} inset which shows the combined emergent spectrum in the near infrared). The squares show the MIPS, IRAC, 2MASS ({\it red}), and IRSF ({\it black}) photometry. The {\it red} and {\it black} MIPS/IRAC photometry correspond to SAGE epoch 1 and 2 data respectively. The {\it red} curves show the near infrared and SAGE--Spec (IRS) spectra. The MIPS, IRAC, near infrared photometry (2MASS and IRSF), near infrared spectrum and IRS spectrum were all obtained at different times, and thus do not uniquely constrain the models. The best matches to object 142 epoch 1 photometry, 2MASS photometry, and the IRS and near infrared spectra result from models with optical depth near 0.45 in the SiC 11.3 \mic \ feature (Figure {\it upper} inset). The IRS spectrum has been shifted slightly to match the 24 \mic \ SAGE epoch 1 photometry and the near infrared spectrum was set to match the 2MASS $K_s$ photometry (2.159 \mic). See text.
\label{2dustb}}
\end{figure}

In Figure~\ref{2dustb}, we show models compared to the IRAC and MIPS epoch 1 photometry (the more reddened epoch). The SAGE--Spec IRS spectrum is also shown as is the near infrared spectrum and photometry.  The SAGE--Spec fluxes and MIPS photometry are all relatively similar. We scaled the IRS spectrum by a factor of 1.07 to match the epoch 1 MIPS 24 \mic \ photometry.
With this correction, the IRS spectrum is a good match to the IRAC epoch 1 photometry at 6 and 8 \mic \ as well. Individual models were scaled to match the 24 \mic \ flux and 11.3 \mic \ SiC feature by eye. 

The best match occurs for an optical depth of 0.45 in the SiC 11.3 \mic \ feature; this model also matches the $K-$band and $H-$band photometry for 2MASS. No models do a good job of matching the near infrared, IRS spectral region {\it and} the shortest IRAC photometry point at 3.6 \mic. But the 0.45 optical depth model does a good job at matching the epoch 1 {\it Spitzer} photometry, IRS spectrum, the 2MASS photometry, and the morphology of the near infrared spectrum (strong veiling and slope). The resulting {\em dust} mass-loss rates are a few times $10^{-9}$ M$_\odot$ yr$^{-1}$. 

The near infrared spectrum was scaled to the 2MASS data point in Figure~\ref{2dustb}. The actual spectrum clearly exhibits more veiling than the 0.45 optical depth model. This suggests that the near infrared spectrum was taken during an even lower brightness phase of the cycle for object 142 than the 2MASS and IRAC epoch 1 photometry. A model with optical depth of 0.5 (in the SiC feature) is shown in the inset and results in even more veiling, but lower near and mid infrared fluxes. Despite this inconsistency which results from the non-simultaneous observations, the models appear to give a plausible explanation of the XAGB stars.

%-----------------------------------------------

The models are consistent with the picture that the XAGB stars, including those with the reddest colors, are mainly C--rich objects on the AGB with varying amounts of dust production in their circumstellar environments. The near infrared and mid infrared spectra of XAGB stars are well matched by C--star core with a dust shell exhibiting modest mass loss. The SiC emission line at 11.3 \mic \ is well matched by the model, but there are several absorption features in the near and mid infrared that are not included in the model and hence not reproduced. The nature of the absorption at 2.26 \mic \ remains unknown because our models do not explicitly include the stellar atmosphere beyond a low resolution input SED. There are other absorptions in the IRS spectrum of XAGB stars. \citet{srini10} showed how these could arise in the circumstellar envelope from warm gas near the inner dust radius. Detailed modeling of the molecular absorptions will be a subject of future work requiring more sophisticated models and perhaps higher resolution near infrared spectra.

The reader is reminded that objects become C--rich when sufficient carbon--rich material is dredged up as the star evolves along the AGB to produce a C/O ratio greater than unity \citep{ir83} at the surface. The SAGE data set spans all ages and masses in the LMC central region. The bulk of the objects are old and low mass, hence they form the main RGB--OAGB--CAGB-XAGB sequence (the latter step inferred from this study and the arguments above). This is a $\sim$ monotonic sequence in luminosity as expected for a more or less uniformly old population ("single age"). There are more luminous O--rich sources than the XAGB stars, however. This is a result of their younger age and high mass. They evolve quickly before the carbin--rich material can be dredged up to the surface.

Object 94 may be an example of an XAGB C--star with maximal veiling such that even the CO band heads and 2.26 \mic \ feature are no longer present at the signal-to-noise of our spectrum. Unfortunately, the IRS spectrum for this object was not obtained (an error was made on positioning and another target was observed in its place). Thus we have no mid infrared spectrum to further characterize this source. 

\subsection{O--rich Sources in the Extreme AGB}

Four objects (7, 81, 88, and 90) whose colors and magnitudes place them on or near the XAGB are classified as O--rich PAGB stars by\citet{wood11}; see also \citet{gruendlChu09} for object 7. The PAGB stage is an extremely short phase near the end of a low or intermediate mass star (\aple 8 M$_{\odot}$) lifetime \citep[see the review by][]{vanwinckel03}. This phase sees the bulk of the envelope of the star expanding away from the stellar $``$core$''$, but before the core ionizes the (detached) envelope as a planetary nebula (PN). The core is expected to show the results of nuclear processing in its atmosphere including $s-$process elements. Depending on the mass of the star and its evolution above the giant branch, it may produce a C--rich or O--rich PAGB envelope (i.e. with or without a third dredge up, respectively). The C--rich or O--rich chemistry is evident in the mid infrared spectra of these stars, and this is in part used to determine the PAGB nature \citep{wood11}. The shaping of the imminent PN also begins in this phase, and a circumstellar disk may play an important role \citep{balickFrank02}. A detailed look at PAGB stars with {\it Spitzer} IRS spectra is given by \citet{matsuura13}.

While both C--rich and O--rich PAGB stars were observed by {\it Spitzer} as part of the SAGE--Spec sample \citep{wood11}, only O--rich PAGB were included in the present sample (the C--rich objects in the SAGE--Spec sample were below the brightness cut--off for the sample observed with SOAR/OSIRIS). Two of the four PAGB stars (objects 88 and 90) are classified as RV Tau variables (a subset of PAGB stars) based upon their MACHO light curves \citep{alcock98}.

\begin{figure}
\plotone{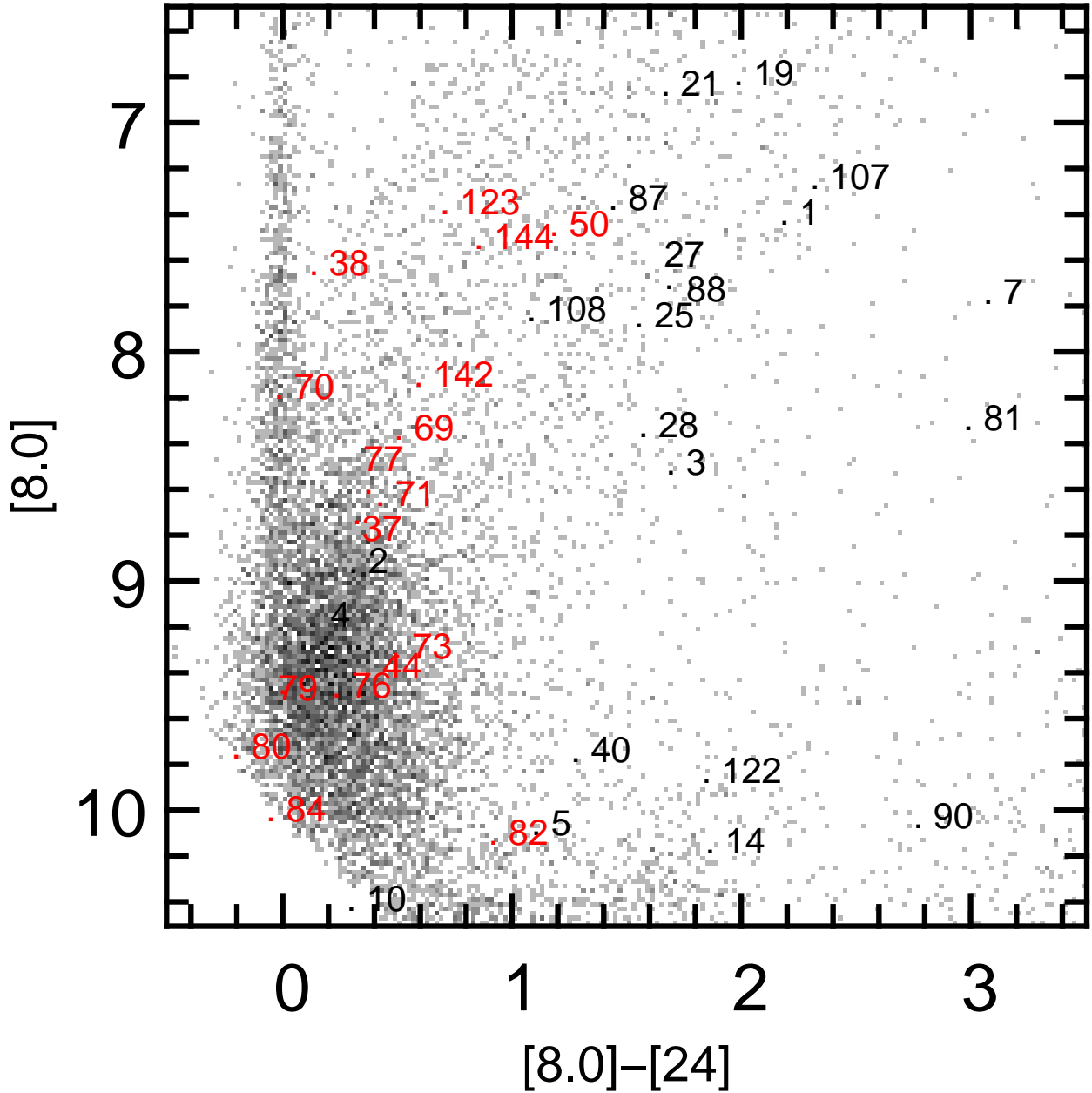} 
\caption{[8.0]$-$[24] $vs.$ [8.0] color--magnitude diagram (CMD) from the SAGE Survey \citep{meixner06}. The SAGE--Spec sub--sample of objects observed here are labeled as {\it black} for O--rich and {\it red} for C--rich stars. The density of objects has a maximum of 11 sources per 0.02 mag $\times$ 0.02 mag bin. This figure appears in color in the online version of the paper.
\label{cmd24}}
\end{figure}

Object 7 appears to have a normal M--type near infrared spectrum typical of a bright O--rich AGB star or M supergiant (see Figure~\ref{other}). Indeed the location of this source is at the top or above the main XAGB locus consistent with an extension to redder colors of the bright O--rich AGB stars in Figure~\ref{cmd}. Object 7 actually lies on a sequence of dusty mass loss stars noted by \citet{blum06}; see their Figure~11. The analogous sequence is shown in the [8.0]$-$[24] $vs.$ [8.0] CMD (Figure~\ref{cmd24}) and is populated by stars 15, 14, 122, 81, and 7 (and possibly 90). 

Object 88 exhibits emission in the CO band head. CO band head emission has been associated with very dense and hot gas in a circumstellar disks in both young stellar objects and evolved stars \citep{sco83, mcgregor88}. Circumstellar disks are also present in (at least) some PAGB stars \citep{sahai99}. \citet{gielen09} presented a detailed study of object 88, also known as MACHO 79.5501.13. The object exhibits a double peaked SED with a strong near infrared excess in the $K$ band; this SED is typical of a class of RV Tau stars with disks. The CO emission (see Figure~\ref{other}) thus corroborates this picture nicely, though as far as we know there are no other O--rich PAGB stars with near infrared spectra showing such a disk signature. Object 90, also an RV Tau star, has a spectrum most similar to O--rich stars (Figures~\ref{orich1} and \ref{orich2}). However there is no strong Ca or Na exhibited in the spectrum. This may be a consequence of the lower signal--to--noise of our spectrum. Object 90 lies in a position below the normal C--star locus in Figure~\ref{cmd}.

Object 81 also appears to be an O--rich star based upon its near infrared spectrum. It is very similar in appearance to the spectrum of object 90, but is not a known RV Tau star.  

\subsection{Mass Loss on the XAGB}

As noted by \citet{kemper10}, it appears that most XAGB stars are C--rich. \citet{zijlstra06} observed a sample of XAGB stars selected based upon their 2MASS colors. All but two turned out to be C--rich. To our knowledge, this is the most unbiased selection of a large sample of stars for follow--up spectroscopy to date. However, the \citet{zijlstra06} sample is only $\sim$ 30 objects compared to the 1100 XAGB stars in the central 7$^{\circ} \times 7^{\circ}$ of the LMC \citep[][i.e., 3$\%$]{blum06}. Hundreds ($\sim$ 300) of mid infrared spectra have been obtained in the LMC for objects in the XAGB region, but they are biased towards brighter magnitudes \citep{kastner08, buchanan09} and/or were selected based on prior information to study targeted classes of objects \citep{sloan08, kemper10}. The large numbers of spectra covering most of the XAGB argue that most objects are indeed C--rich, especially if young stellar objects are excluded based on additional imaging and wavelength data \citep{gruendlChu09}. This is further supported by analysis of multi wavelength photometry. \citet{riebel12} fit the photometry of the entire AGB population in the LMC and found that 97 $\%$ of their 1340 XAGB candidates were C--rich. The SED-based classification has good agreement ($>$80$\%$) spectroscopic identification. Still, an unbiased spectroscopic assessment of the XAGB stellar content has not been done. 
See Figure~14 of \citet{wood11} for a recent $J-$[8.0] $vs.$ [8.0] CMD with the combined sample of mid infrared spectroscopic sources over--plotted. 

Figure~\ref{cmd24} shows the SAGE objects in the [8.0]$-$[24] $vs.$ [24] color--magnitude plane. The striking difference between this plot and Figure~\ref{cmd} is that the O--rich stars are redder in this new diagram where as the C--rich objects are generally redder in the other (for a given [8.0] magnitude. This is understood for the C--rich stars in $J-$[8.0] $vs.$ [8.0] diagram because their SEDs are such that the $J$ flux is on the Wien side of the black body curve, while the [8.0] flux is on the Raleigh--Jeans side. On the other hand, the O--rich stars have SEDs with $J-$band fluxes near the peak of the distribution.

\citet{srini10} made a detailed mass loss calculation for object 123 and \citet{sargent10} did a similar analysis for two O--rich sources, one bright AGB star (HV5715) and one fainter AGB star (SSTISAGEMC J052206.92--715017.7). We did not observe either of the two O--rich stars analyzed by \citet{sargent10}; however, objects 2 and 3 of our sample are similar to HV5715 and object 122 is similar to SSTISAGEMC J052206.92--715017.7. The O--rich stars in our sample similar in colors and magnitudes to HV5715 and SSTISAGEMC J052206.92--715017.7 appear to be normal late type stars with O--rich near infrared spectra. Object 123 analyzed by \citet{srini10} is an XAGB star as discussed above. 

Perhaps remarkably, the dust mass loss rates computed for these three objects and object 142 discussed above which span a large region in the CMD are quite similar. The values range from 2 to 3 $\times$ 10$^{-9}$ M$_{\odot}$ yr$^{-1}$. The XAGB stars, object 123 and 142 (C--rich) and SSTISAGEMC J052206.92--715017.7 are relatively low mass. The former are approximately 2 M$_{\odot}$ and the latter 3 M$_{\odot}$ based upon position in the Hertzprung--Russell diagram. See\citet[see][]{srini10} and \citet{sargent10} for details. HV5715 is more massive, approximately 7 M$_{\odot}$ \citep{sargent10}. SSTISAGEMC J052206.92--715017.7 sits on the reddest tip of of a feature attributed to high mass loss stars in the [8.0]$-$[24] $vs.$ [24] CMD presented by \citet{blum06}, so it may not be surprising that a relatively low mass $``$normal$''$ O--rich AGB star has similar mass loss as an XAGB (C--rich) star. 

\section{Summary}

We have presented a sample of 40 SAGE--Spec {\it Spitzer} IRS evolved stars in the Large Magellanic Cloud. These objects span the bright end of the color--magnitude diagram (CMD) from the tip of the normal asymptotic giant branch (AGB) to the top of the red supergiant (RSG) branch to the reddest colors on the extreme AGB (XAGB) as defined in the $J-$[8.0] $vs.$ [8.0] CMD.

The near infrared spectra are an excellent match to the characterization provided by the IRS spectral properties. The C--rich spectral morphology owing to numerous CN and CO lines is unmistakable compared to the O--rich atomic and CO features. 

We find the nature of the reddest XAGB stars is best explained by C--rich cores surrounded by dusty mass loss envelopes. We modeled one of these sources (142) with the dust radiative transfer code 2Dust \citep[following][]{srini10} and find plausible models to explain the {\it Spitzer} IRS spectrum, IRAC and MIPS photometry as well as the observed near infrared spectral morphology. This latter result stems from adding normal C--star spectra to the 2Dust input SED and then comparing the resulting emergent spectrum which includes the effects of the overlying envelope emission and extinction to the observed near infrared spectrum. These quantitative results are in agreement with conclusions made for less obscured C--stars \citep{fh72} based on the anti--correlation of CO absorption strength and infrared color. We conclude that modest mass loss rates ($\sim$ few by 10$^{-9} $ M$_{\odot}$~yr$^{-1}$) reproduce both the near infrared and mid infrared properties on the XAGB. 

Four of the present sample of objects were selected as post AGB stars, a short lived phase of stellar evolution between the heavy mass loss of the AGB phase and the planetary nebula phase. These objects in our sample exhibit more O--rich like near infrared spectra. One of the sample (HV 915, object 88) has CO 2.3\mic \ band head emission. Circumstellar emission is known in some other RV Tau (i.e. a PAGB subclass) stars; to our knowledge, this is the first report of a near infrared spectrum with a disk like circumstellar signature for a RV Tau star. This emission and a disk origin are consistent with previous work showing the spectral energy distribution of object 88 is best explained by a disk.

The O--rich PAGB stars are overrepresented in our sample because \citet{kemper10} selected the parent sample based on properties of objects they wanted to investigate, not an unbiased sample of objects  on the XAGB. Nevertheless, the picture that emerges \citep[see][]{zijlstra06, kastner08, sloan08, wood11} is that most of the XAGB is comprised of otherwise normal C--rich stars (typically a few solar masses) that have produced dusty shells due to mass loss. The O--rich PAGB stars and other sources (planetary nebulae,  HII regions, young stellar objects, etc) projected onto the XAGB are interlopers or short lived, unrelated, phases of stellar evolution. 

The authors would like to acknowledge Ken Hinkle (NOAO) for useful discussions about CN lines in the near infrared spectrum of evolved stars. The authors thank Masaaki Otsuka (ASIAA) for his assistance to B. Ling with near infrared spectral reductions. FK acknowledges generous support from the Taiwan National Science Council, grant number: NSC100-2112-M-001-023-MY3. The authors thank Jay Frogel 
for his careful reading of the manuscript as referee and his comments and 
suggestions which have improved our paper. 

%--------------------------References------------------------------------

\bibliographystyle{apj.bst}
\bibliography{ms}

%----------------------------------FIGURES--------------------------------

%---------------------------------------------Table 1----------------------------------------
% Target list

\newcounter{mytabnum}
%\addtocounter{mytabnum}{1}
\refstepcounter{mytabnum}

\begin{deluxetable}{lcccccccccc}
\rotate
\label{point}
\tabletypesize{\footnotesize}
\tablecolumns{11} 
\tablewidth{0pt}
\label{targets}
\tablecaption{{\it SAGE--Spec Targets}}
\tablehead{
\colhead{Object} &
\colhead{$K$} &
\colhead{$J-$[8.0]} &
\colhead{[8.0]} &
\colhead{[8.0]--[24]} &
\colhead{[24]} &
\colhead{SSID\tablenotemark{a}} &
\colhead{Other Name\tablenotemark{b}}&
\colhead{CMD Type\tablenotemark{c}} &
\colhead{NIR\tablenotemark{d}} &
\colhead{PSC\tablenotemark{e}}
}
\startdata 
28   & 8.83 & 1.50 & 8.37 & 1.55 & 6.82 & 129 &  NGC 2004 Wes 6-14 & RSG & O star & RSG \\
108 & 8.41 & 1.55 & 7.87 & 1.06 & 6.80 & 122 & HV 5879  & RSG & O star & RSG \\
40   & 10.23 & 1.71 & 9.79 & 1.26 & 8.54 & 143 &  SSTISAGEMC J053343.98--705901.9 & B O* & O star & OAGB \\
4     & 9.88 & 1.74 & 9.42 & 0.21 & 9.21 & 178 & SSTISAGEMC J054314.12-703835.1 & B O* & O star & OAGB \\
5     & 10.68 & 1.79 & 10.11 & 1.08 & 9.02 & 197 & SSTISAGEMC J060053.62--680038.8 & B O* & O star & OAGB \\
10   & 10.99 & 1.83 & 10.44 & 0.28 & 10.16 & 13 &  SSTISAGEMC J045309.39--681710.8& O* & O star & OAGB \\
2     & 9.52 & 1.84 & 8.96 & 0.29 & 8.68 & 173 & WHO G 494 & B O* & O star & OAGB \\
15   & 11.28 & 1.91 & 10.62 & 1.40 & 9.22 & 89 & SSTISAGEMC J052014.24--702931.0 & N O* & O star & OAGB \\
25   & 8.74 & 1.95 & 7.89 & 1.53 & 6.36 & 170 & NGC 2100 Robb 4 & RSG & O star & RSG \\
19   & 7.74 & 2.10 & 6.84 & 1.97 & 4.87 & 27 & MSX LMC 1271  & RSG & O star & RSG \\
14   & 11.08 & 2.17 & 10.19 & 1.84 & 8.35 & 91 & SSTISAGEMC J052051.83--693407.6 & O* & O star & OAGB \\
122 & 10.87 & 2.23 & 9.88 & 1.83 & 8.05 & 159 & SSTISAGEMC J053945.40--665809.4 &O* & O star& OAGB \\
107 & 8.34 & 2.25 & 7.29 & 2.30 & 4.99 & 147 & HV 2700  & RSG & O star & RSG \\
3     & 9.53 & 2.28 & 8.54 & 1.67 & 6.87 & 61 & SSTISAGEMC J051059.07--685613.7 &O* & O star & OAGB \\
27  & 8.82 & 2.30 & 7.73 & 1.66 & 6.06 & 172 & W61 6-57 & RSG & O star & RSG \\
21  & 8.05 & 2.38 & 6.88 & 1.65 & 5.23 & 135 & NGC 2011 SAGE IRS 1 & RSG & O star & RSG \\
82  & 11.13 & 2.44 & 10.15 & 0.90 & 9.26 & 132 & KDM 4665 & C* & C star  & CAGB\\
73  & 10.40 & 2.51 & 9.34 & 0.48 & 8.86 & 120 & OGLE 052825.96-694647.4 & C* & C star & CAGB \\
76  & 10.45 & 2.66 & 9.51 & 0.21 & 9.29 & 136 & KDM 4718 & C* & C star  & CAGB \\
1    & 8.78 & 2.69 & 7.44 & 2.17 & 5.28 & 165 & MSX LMC 947 & O* & O star & OAGB \\
44  & 10.43 & 2.84 & 9.43 & 0.35 & 9.08 & 83 & SSTISAGEMC J051832.64--692525.5 & C* & C star & CAGB \\
79  & 10.69 & 2.91 & 9.49 & -0.02 & 9.51 & 23 & KDM 1238 & C* & C star & CAGB \\
84  & 11.42 & 2.94 & 10.05 & -0.07 & 10.12 & 57 & KDM 2187 & C* & C star  & CAGB\\
71  & 9.95 & 3.00 & 8.67 & 0.40 & 8.27 & 33 & LMC BM 12-14 & C* & C star  & CAGB \\
80  & 11.01 & 3.10 & 9.78 & -0.22 & 10.00 & 49 &  SSTISAGEMC J050629.61--685534.9& C* & C star  & CAGB \\
69  & 9.69 & 3.10 & 8.38 & 0.49 & 7.90 & 31 & KDM 1691 & B C* & C star & CAGB \\
90  & 12.18 & 3.16 & 10.08 & 2.75 & 7.33 & 157 & HV 12631 & Below C* & O rich----\tablenotemark{e} & RV Tau  \\
37  & 10.03 & 3.27 & 8.75 & 0.30 & 8.45 & 12 & KDM 764 & C* & C star & CAGB \\
70  & 9.78 & 3.39 & 8.21 & -0.04 & 8.24 & 30 & KDM 1656 & B C* & C star & CAGB\\
77  & 10.49 & 4.02 & 8.62 & 0.35 & 8.27 & 45 &  SSTISAGEMC J050607.50--714148.4& E XAGB & C star, CN?& CAGB \\
7   & 10.82 & 4.11 & 7.79 & 3.45 & 4.34 & 177 & SSTISAGE1C J054310.86--672728.0 & Above XAGB & O star & OPAGB \\
81 & 11.09 & 4.12 & 8.34 & 2.97 & 5.37 & 56 &  SSTISAGEMC J050830.51--692237.4& E XAGB & O rich?\tablenotemark{e} & OPAGB \\
38 & 10.06 & 4.97 & 7.67 & 0.12 & 7.55 & 51 & SHV 0507252--690238 & Mid XAGB & C star, CN? & CAGB\\
94 & 9.91 & 5.23 & 7.24 & 0.45 & 6.79 & 145 & wrong star in IRS slit & Mid XAGB & C star & CAGB \\
88 & 12.09 & 5.36 & 7.73 & 1.76 & 5.97 & 73 & HV 915 & Mid XAGB & CO Emission & O* RV Tau \\
87 & 11.89 & 5.98 & 7.38 & 1.42 & 5.96 & 94 & HV942 & L XAGB & no source detected & R CrB \\
170 & 14.16 & 6.02 & 9.87 & 3.06 & 6.81 & 193 & SSTISAGEMC J055143.27--684543.0  & galaxy & featureless & galaxy \\
123 & 10.89 & 6.76 & 7.40 & 0.69 & 6.72 & 66 & OGLE J051306.52--690946.4 & L XAGB & C star, "featureless", CN? & CAGB \\
142 & 11.78 & 6.97 & 8.15 & 0.57 & 7.58 & 103 & SSTISAGEMC J052405.31--681802.5 & L XAGB & C star, "featureless" CN? & CAGB \\
50 & 11.35 & 7.86 & 7.49 & 1.16 & 6.33 & 60 &  SSTISAGEMC J051028.27--684431.2& L XAGB & C star, "featureless" & CAGB \\
144 & 11.81 & 8.24 & 7.55 & 0.83 & 6.72 & 98 & OGLE J052242.09--691526.2 & L XAGB & C star, "featureless" & CAGB

\enddata

\tablenotetext{a}{SSID is the {\it Spitzer} observation ID given in \citet{kemper10}. See also \cite{wood11} for details of the object's {\it Spitzer} IRS spectrum and classification.}

\tablenotetext{b}{Other names adopted from \citet{wood11}.}

\tablenotetext{c}{The expected object type based upon the position in the $J-$[8.0] $vs.$ [8.0] color--magnitude diagram (CMD); see Figure~\ref{cmd}. The type of object for a given position in the CMD is defined in \citet{blum06}. O* $=$ O--rich asymptotic giant branch (AGB). RSG $=$ red supergiant. B O* $=$ bright AGB (between the AGB and RSG sequences). B C* $=$ bright C--star; XAGB $=$ extreme AGB. E, Mid, and L $=$ early (or blue), middle and late (or red) XAGB which roughly divides the XAGB sequence in three sections. }

\tablenotetext{d}{Object type based upon morphology of the near infrared spectra presented herein. See text and Figures~\ref{orich1}, \ref{orich2}, \ref{crich}, \ref{x}, and \ref{other}. CN? $=$ possible CN feature seen only in the XAGB stars; see text.}

\tablenotetext{e}{Point Source Catalog type or characteristics based on {\it Spitzer} IRS spectra \citep{wood11}.}

\tablenotetext{f}{Objects 90 and 81 appear to have a O--rich near infrared spectrum, but the signal to noise is not high (see Figure~\ref{other}) and \S 5.2.}

\end{deluxetable}

%--------------------------------------------------------------------------
\end{document}